\documentstyle[twoside,psfig,hyperon99_paper]{article}
\title{Hyperon Physics---a Personal Overview}
\author{Barry R. Holstein\\
Department of Physics and Astronomy\\
University of Massachusetts\\
Amherst, MA  01003}
\begin{document}
\begin{abstract}
A range of issues in the field of hyperon physics is presented, together
with an assessment of where important challenges remain. 
\end{abstract}
\maketitle
\section{Introduction}
The subject of hyperon physics is a vast one, as indicated by the fact that
this workshop will run for three days, with presentations involving a 
range of different issues.  Obviously it would be impossible for me to
cover all of the interesting features in this introductory presentation.
Instead, I will present a very personal picture of some of the issues in 
hyperon physics which {\it I} think need to be answered, and will trust the
various speakers to fill in areas which I have omitted.

\section{Hyperon Processes}
I have divided my presentation into sections, which cover the various arenas
which I think need attention:
\subsection{Nonleptonic Hyperon Decay}
The dominant decay mode of the ${1\over 2}^+$ hyperons is, of course, 
the pionic decay $B\rightarrow B'\pi$.  On the theoretical side there
remain two interesting and important issues which have been with us since the
1960's---the origin of the $\Delta I=1/2$ rule and the S/P-wave 
problem\cite{dghbk}:
\begin{itemize}

\item [i)]  The
former is the feature that $\Delta I=3/2$ amplitudes are suppressed with
respect to their $\Delta I=1/2$ counterparts by factors of the order of 
twenty or so.  This suppression exists in both hyperon as well as kaon
nonleptonic decay and, despite a great deal of theoretical work, there
is still no simple explanation for its existence.  The lowest order
weak nonleptonic $\Delta S=1$ Hamiltonian possesses comparable $\Delta
I=1/2$ and $\Delta I=3/2$ components and leading log gluonic effects
can bring about a $\Delta I=1/2$ enhancement of a factor of three to
four or so\cite{llog}.  The remaining factor of five seems to arise
from the validity of what is called the Pati-Woo theorem in the baryon
sector\cite{pw} while for kaons it appears to be associated with detailed
dynamical structure\cite{ddyn}.  Interestingly the
one piece of possible evidence for its violation comes from a hyperon
reaction----hypernuclear decay\cite{hnuc}.  A hypernucleus is produced when a 
neutron in an atomic nucleus is replaced by a $\Lambda$.  In this case
the usual pionic decay mode is Pauli suppressed, and the 
hypernucleus primarily
decays via the non-mesonic processes $\Lambda p\rightarrow np$ and
$\Lambda n\rightarrow nn$.  There does exist a rather preliminary
indication here of a possibly significant $\Delta I=1/2$ rule
violation, but this has no Fermilab relevance and will have to be
settled at other laboratories\cite{dub}.  

\item [ii)]  The latter problem is not as well known but has been a 
longstanding difficulty to those of us theorists who try to calculate these
things.  Writing the general decay amplitude as
\begin{equation}
{\rm Amp}=\bar{u}(p')(A+B\gamma_5)u(p)
\end{equation}  
The standard approach to such decays goes back to current algebra
days and expresses the S-wave (parity-violating) amplitude---$A$---as
a contact term---the baryon-baryon matrix element of the 
axial-charge-weak Hamiltonian commutator.  
The corresponding P-wave (parity-conserving) 
amplitude---$B$---uses a simple pole model ({\it cf.} Figure 1) 
with the the weak baryon to baryon
matrix element given by a fit to the S-wave sector.  Parity violating 
$BB'$ matrix elements are neglected in accord with the Lee-Swift 
theorem\cite{ls}.  
\begin{figure}
\centerline{\psfig{figure=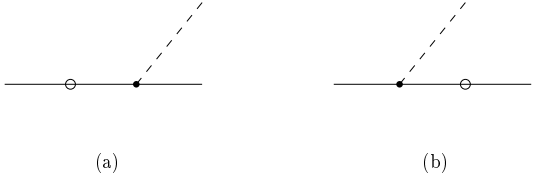,width=7.0cm}}
\caption{Pole diagrams used to calculated parity-conserving nonleptonic
hyperon decay.}
\end{figure}
With this procedure
one can obtain a good S-wave fit but finds P-wave amplitudes which are 
in very poor agreement with experiment.  On the other hand, one can fit
the P-waves, in which case the S-wave predictions are very bad\cite{dghrv}.  
Clearly
the solution requires the input of additional physics, such as inclusion
of $(70,1^-)$ intermediate states as done in an SU(6) calculation 
by Le Youaunc et al.\cite{you} or of intermediate ${1\over 2}^-$ 
and ${1\over 2}^+$
resonant states by Borasoy and myself in a chiral picture\cite{bh1}.  
\end{itemize}

\noindent In either case, we do {\it not} require more and better data.  
The issues are already clear.  What we need is more and better theory!

Where we {\it do} need data involves 
the possibility of testing the standard model
prediction of CP violation, which predicts the presence of various asymmetries
in the comparision of hyperon and antihyperon nonleptonic decays\cite{dcp}.  
The basic
idea is that one can write the decay amplitudes in the form
\begin{equation}
A=|A|\exp i(\delta_S+\phi_S),\quad B=|B|\exp i(\delta_P+i\phi_P)
\end{equation} 
where $\delta_S,\delta_P$ are the strong S,P-wave phase shifts at the 
decay energy of the mode being considered and $\phi_S,\phi_P$ are CP-violating
phases which are expected to be of order $10^{-4}$ or so in standard model
CP-violation.  One can detect such phases by comparing hyperon and antihyperon
decay parameters.  Unfortunately nature is somewhat perverse here in that
the larger the size of the expected effect, the more difficult the 
experiment.  For example, the asymmetry in the overall decay rate, which is
the easiest to measure, has the form\footnote{Here $B^r$ indicates a reduced
amplitude---$B^r= B(E'-M_{B'})/(E'+M_{B'})$.}
\begin{eqnarray}
C&=&{\Gamma-\bar{\Gamma}\over \Gamma+\bar{\Gamma}}\nonumber\\
&\sim& 
\left(-2(A_1A_3\sin(\delta_S^1
-\delta_S^3)\sin(\phi_S^1-\phi_S^3)\right.\nonumber\\
&+&\left.B_1^rB_3^r\sin(\delta_P^1-\delta_P^3)\sin(
\phi_P^1-\phi_P^3)\right)\nonumber\\
&/& |A_1|^2+|B_1^r|^2
\end{eqnarray}
where the subscripts, superscripts 1,3 
indicate the $\Delta I={1\over 2},{3\over 2}$ 
component of the amplitude.  We see then that there is indeed sensitivity to
the CP-violating phases but that it is multiplicatively 
suppressed by {\it both} the the 
strong interaction phases ($\delta\sim 0.1$) as well as by the $\Delta I=
{3\over 2}$ suppression $A_3/A_1\sim B_3/B_1\sim 1/20$.  Thus we 
find $C\sim\phi/100\sim 10^{-6}$
which is much too small to expect to measure in present generation experiments.

More sanguine, but still not optimal, is a comparison of the asymmetry
parameters $\alpha$, defined via
\begin{equation}
W(\theta)\sim1+\alpha\vec{P}_B\cdot\hat{p}_{B'}
\end{equation}
In this case, one finds
\begin{eqnarray}
A&=&{\alpha+\bar{\alpha}\over \alpha-\bar{\alpha}}=-\sin(\phi_S^1-\phi_P^1)
\sin(\delta_S^1-\delta_P^1)\nonumber\\
&\sim& 0.1\phi\sim 4\times 10^{-4}
\end{eqnarray} 
which is still extremely challenging.

Finally, the largest signal can be found in the combination
\begin{equation}
B={\beta+\bar{\beta}\over \beta-\bar{\beta}}=\cot(\delta_S^1
-\delta_P^1)\sin(\phi_S^1-\phi_P^1)\sim\phi
\end{equation}
Here, however, the parameter $\beta$ is defined via the general expression
for the final state baryon polarization
\begin{eqnarray}
<\vec{P}_{B'}>&=&{1\over W(\theta)}\left((\alpha+\vec{P}_B\cdot\hat{p}_{B'})
\hat{p}_{B'}\right.\nonumber\\
&+&\left.\beta\vec{P}_B\times\hat{p}_{B'}
+\gamma(\hat{p}_{B'}
\times(\vec{P}_B\times\hat{p}_{B'}))\right)\nonumber\\
\quad
\end{eqnarray}
and, although the size of the effect is largest---$B\sim 10^{-3}$---this
measurement seems out of the question experimentally.  

Despite the small size of these effects, the connection with standard model
CP violation and the possibility of finding larger effects due to new
physics demands a no-holds-barred effort to measure these parameters.

\subsection{Nonleptonic Radiative Decay}
Another longstanding thorn in the side of theorists attempting to understand
weak decays of hyperons is the nonleptonic radiative mode $B\rightarrow
B'\gamma$\cite{zrev}.  In this case one can write 
the most general decay amplitude
as 
\begin{eqnarray}
{\rm Amp}&=&{e\over M_B+M_{B'}}\epsilon^{*\mu}q^\nu\nonumber\\
&\times&\bar{u}(p')\left(-i\sigma_{\mu\nu}C
-i\sigma_{\mu\nu}\gamma_5D)\right)u(p)
\end{eqnarray}
where $C$ is the magnetic dipole (parity conserving) amplitude and $D$ is its
(parity conserving) electric dipole counterpart.  There are two 
quantities of interest in the
analysis of such decays---the decay rate and photon asymmetry, which go as
\begin{equation}
\Gamma\sim|C|^2+|D|^2,\quad A_\gamma={2{\rm Re}C^*D\over |C|^2+|D|^2}
\end{equation}
The difficulty here is associated with ``Hara's Theorem'' which requires that
in the SU(3) limit the parity violating decay amplitude must vanish for decay
between states of a common U-spin multiplet---{\it i.e.} $\Sigma^+\rightarrow
p\gamma$ and $\Xi\rightarrow \Sigma^-\gamma$\cite{hara}.  (The proof
here is very much analogous to the 
one which requires the vanishing of the axial tensor form factor in 
nuclear beta decay between members of a common isotopic spin 
multiplet\cite{ht}.)
Since one does not expect significant 
SU(3) breaking effects, we anticipate 
a relatively small photon asymmety parameter for such decays.  However, 
in the case of $\Sigma^+\rightarrow p\gamma$ the asymmetry is known to be
large and negative\cite{pdg}
\begin{equation}
A_\gamma(\Sigma^+\rightarrow p\gamma)=-0.76\pm 0.08\label{eq:aa}
\end{equation} 
and for thirty years theorists have been struggling to explain this result.
In leading order the amplitude is given by the simple pole diagrams, with
the weak baryon-baryon matrix elements being those determined in the 
nonradiative decay analysis.  
\begin{figure}
\centerline{\psfig{figure=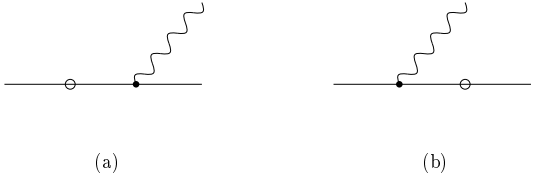,width=7cm}}
\caption{Pole diagrams used to calculate radiative hyperon decay.}
\end{figure}
The Lee-Swift theorem asserts that such matrix
elements must be purely parity conserving in the SU(3) limit and this is
the origin of Hara's theorem in such a model\cite{ls}.  Although SU(3) breaking
corrections have been calculated, none is large enough to explain 
the experimental result---Eq. \ref{eq:aa}\cite{gh}.  
As in the case of the S/P-wave
puzzle, what is clearly required is the inclusion of additional physics
and here too the inclusion of $(70,1^-)$ states by Le Youaunc et al.\cite{you} 
or of
${1\over 2}^-$ and ${1\over 2}^+$ resonant states in a chiral framework by
Borasoy and myself\cite{bh2} 
appears to naturally predict a large negative asymmetry.
However, in order to confirm the validity of these or any model what 
will be required is a set of measurements of {\it both} rates and 
asymmetries for
such decays.  In this regard, it should be noted that theoretically one
expects all asymmetries to be negative in any realistic
model\cite{pz}.  It would be very difficult
to accomodate a large positive asymmetry.  Thus the present particle data
group listing\cite{pdg} 
\begin{equation}
A_\gamma(\Xi^0\rightarrow\Lambda\gamma)=+0.43\pm0.44
\end{equation}      
deserves to be carefully remeasured.

\subsection{Hyperon Beta Decay}
A mode that theory does well in predicting (in fact some would say 
{\it too} well) is that of hyperon beta decay---$B\rightarrow B'\ell\nu_\ell$,
where $\ell$ is either an electron or a muon.  Since this is a semileptonic
weak interaction, the decays are described in general by matrix elements of
the weak vector, axial-vector currents
\begin{eqnarray}
<B'|V_\mu|B>&=&\bar{u}(p')(f_1\gamma_\mu+{-if_2\over M_B+M_{B'}}
\sigma_{\mu\nu}q^\nu\nonumber\\
&+&{f_3\over M_B+M_{B'}}q_\mu)u(p)\nonumber\\
<B'|A_\mu|B>&=&\bar{u}(p')(g_1\gamma_\mu+{-ig_2\over M_B+M_{B'}}
\sigma_{\mu\nu}q^\nu\nonumber\\
&+&{g_3\over M_B+M_{B'}}q_\mu )\gamma_5u(p)\nonumber\\
\quad
\end{eqnarray}
Here the dominant terms are the vector, axial couplings $f_1,g_1$ and the
standard approach is simple Cabibbo theory, wherein one fits the $g_1$
in terms of SU(3) F,D coefficients and $f_1$ using
CVC and simple F coupling.  When this is done, one finds in general a 
very satisfactory fit---$\chi^2/d.o.f\sim 2.0$---which can be made even better 
by inclusion of simple quark model SU(3) breaking effects---$\chi^2/d.o.f.
\sim 0.85$\cite{dhk}.  An output of such a fit is the value of the KM mixing parameter
$V_{us}=0.220\pm 0.003$, which is in good agreement with the value
$V_{us}=0.2196\pm 0.0023$ measured in $K_{e3}$ decay.  
However, differing assumptions about SU(3) 
breaking will lead to slightly modified values.  

The importance of such a measurement of $V_{us}$ has to do with its use as
an input to a test of the standard model via the unitarity prediction
\begin{equation}
|V_{ud}|^2+|V_{us}|^2+|V_{ub}|^2=1
\end{equation} 
From an analysis of B-decay one obtains $|V_{ub}|\sim 0.003$, which when 
squared leads to a negligible contribution to the unitarity sum.  So the
dominant effect comes from $V_{ud}$, which is measured via $0^+-0^+$
superallowed nuclear beta decay---
\begin{equation}
V_{ud}^2={2\pi^3\ln 2m_e^{-5}\over 2G_F^2(1+\Delta_R^V)\bar{\cal F}t}
\end{equation}
Here $\Delta_R^V=2.40\pm 0.08\%$ is the radiative correction and 
$\bar{\cal F}t= 3072.3\pm 0.9$ sec. is the mean (modified) ft-value
for such decays.  Of course, there exist important issues in the analysis
of such ft-values including the importance of isotopic spin breaking
effects and of possible Z-dependence omitted from the radiative corrections, 
but if one takes the above-quoted number as being correct we obtain\cite{htow}
\begin{equation}
\begin{array}{c}
V_{ud}=0.9740\pm0.0005\\
{\rm and}\\
|V_{ud}|^2+|V_{us}|^2+|V_{ub}|^2=
0.9968\pm 0.0014
\end{array}
\end{equation}
which indicates a possible violation of unitarity.  If correct, this would
suggest the existence of non-standard-model physics, but clearly additional
work, both theoretical and experimental, is needed before drawing this
conclusion.

What is needed in the case of hyperon beta decay is good set of data including
rates {\it and} asymmetries, both in order to produce a possibly improved
value of $V_{us}$ but also to study the interesting issue of SU(3) breaking
effects, which {\it must} be present, but whose effects seem somehow
to be hidden.  A related focus of such studies should be the
examination of higher
order---recoil---form factors such as weak magnetism ($f_2$) and the axial
tensor ($g_2$).  In the latter case, Weinberg showed that in the standard
quark model $G=C\exp (i\pi I_2)$-invariance requires $g_2=0$ in neutron
beta decay $n\rightarrow pe^-\bar{\nu}_e$\cite{scc}.  
(This result usually is called
the stricture arising from ``no second class currents.'')  In the SU(3)
limit one can use V-spin invariance to show that $g_2=0$ also obtains for
$\Delta S=1$ hyperon beta decay, but in the real world this condition
will be violated.  A simple quark model calculation suggests that 
$g_2/g_1\sim -0.2$\cite{dh2} but other calculations, such as 
a recent QCD sum rule 
estimate give a larger number---$g_2/g_1\sim -0.5$.
In any case good hyperon beta decay data---with rates and 
asymmetries---will be needed in order to extract the size of such effects.

\subsection{Hyperon Polarizabilities}

Since this subject is not familiar to many physicists, let me spend just
few moments giving a bit of motivation.  The idea goes back to simple classical
physics.  Parity and time reversal invariance, of course, forbid the existence
of a permanent electric dipole moment for an elementary system.  
However, consider the application of a 
uniform electric field to a such a system.
Then the positive charges will move in one
direction and negative charges in the other---{\it i.e.} there will be a
charge separation and an electric dipole moment will be induced.  The size of
the edm will be proportional to the applied field and
the constant of proportionality between the applied field and the induced
dipole moment is the electric polarizability $\alpha_E$
\begin{equation}
\vec{p}=4\pi\alpha_E\vec{E}
\end{equation}
The interaction of this dipole moment with the field leads to an
interaction energy
\begin{equation}
U=-{1\over 2}\vec{p}\cdot\vec{E}=-{1\over 2}4\pi\alpha_E\vec{E}^2,
\end{equation}
where the ``extra'' factor of $1\over 2$ compared to elementary physics
result is due to the feature that the dipole moment is {\it induced}.
Similarly in the presence of an applied magnetizing field $\vec{H}$ there will
be generated an induced magnetic dipole moment
\begin{equation}
\vec{\mu}=4\pi\beta_M\vec{H}
\end{equation}
with interaction energy
\begin{equation}
U=-{1\over 2}\vec{\mu}\cdot\vec{H}=-{1\over 2}4\pi\beta_M\vec{H}^2.
\end{equation}
For wavelengths large compared to the size of the system, the effective
Hamiltonian describing the interaction of a system of charge $e$ and 
mass $m$ with an electromagnetic
field is, of course, given by 
\begin{equation}
H^{(0)}={(\vec{p}-e\vec{A})^2\over 2m}+e\phi,
\end{equation}
and the Compton scattering cross section has the simple Thomson form
\begin{equation}
{d\sigma\over d\Omega}=\left({\alpha_{em}\over m}\right)^2\left({\omega'\over
\omega}\right)^2[{1\over 2}(1+\cos^2\theta)],
\end{equation}
where $\alpha_{em}$ is the fine structure constant and $\omega,\omega'$ 
are the initial, final photon energies respectively.
As the energy increases, however, so does the resolution and
one must also take into account polarizability
effects, whereby the effective Hamiltonian becomes
\begin{equation}
H_{\rm eff}=H^{(0)}-{1\over 2}4\pi(\alpha_E\vec{E}^2+\beta_M\vec{H}^2).
\end{equation}
The Compton scattering cross section from such a system (taken, for simplicity,
to be spinless) is then 
\begin{eqnarray}
{d\sigma\over d\Omega}&=&\left({\alpha_{em}\over m}\right)^2\left({\omega'\over
\omega}\right)^2\left({1\over 2}
(1+\cos^2\theta)\right.\nonumber\\
&-&\left.{m\omega\omega'\over \alpha_{em}}[{1\over
2}(\alpha_E+\beta_M)(1+\cos\theta)^2\right.\nonumber\\
&+&\left.{1\over 2}(\alpha_E-\beta_M)
(1-\cos\theta)^2]\right)\label{eq:sss}
\end{eqnarray}
It is clear from Eq. \ref{eq:sss}
that from careful measurement of the differential scattering cross section,
extraction of these structure dependent polarizability terms is possible
provided 
\begin{itemize}
\item [i)] that the energy is large enough that these terms are 
significant compared to the
leading Thomson piece and 
\item [ii)] that the energy is not so large that higher order
corrections become important.  
\end{itemize}
In this fashion the measurement of electric and
magnetic polarizabilities for the proton has recently been accomplished 
at SAL and at MAMI using
photons in
the energy range 50 MeV  $<\omega <$ 100 MeV, yielding\cite{PPol}
\footnote{Results for the neutron extracted from $n-Pb$ scattering cross
section
measurements have been reported\cite{npol}, 
but have been questioned\cite{ques}.
Extraction via studies using a deuterium target may be possible
in the future\cite{bean}.}
\begin{eqnarray}
\alpha_E^p&=&(12.1\pm 0.8\pm 0.5)\times 10^{-4}\; {\rm fm}^3\nonumber\\
\beta_M^p&=&(2.1\mp 0.8\mp 0.5)\times 10^{-4}\; {\rm fm}^3. \label{abexp}
\end{eqnarray}
Note that in
practice one generally exploits the strictures of causality and unitarity as
manifested
in the validity of the forward scattering dispersion relation, which yields the
Baldin sum rule\cite{bgm}
\begin{eqnarray}
\alpha_E^{p,n}&+&\beta_M^{p,n}={1\over 2\pi^2}\int_0^\infty{d\omega\over
\omega^2}\sigma_{\rm tot}^{p,n}\nonumber\\
&=&\left\{
\begin{array}{ll}(13.69\pm 0.14)\times 10^{-4}{\rm fm}^3& {\rm proton}\\
                 (14.40\pm 0.66)\times 10^{-4}{\rm fm}^3& {\rm neutron}
\end{array}\right.\nonumber\\
\quad
\end{eqnarray}
as a rather precise constraint because of the small uncertainty associated 
with the photoabsorption cross section $\sigma_{\rm tot}^p$.

As to the meaning of such results we can compare with the corresponding 
calculation of the electric polarizability of the hydrogen atom, 
which yields\cite{merz}
\begin{equation}
\alpha_E^H={9\over 2}a_0^2\quad{\rm vs.}\quad\alpha_E^p\sim
10^{-3}<r_p^2>^{3\over 2}
\end{equation}
where $a_0$ is the Bohr radius.  Thus the polarizability of the hydrogen
atom is of order the atomic volume while that of the proton is 
only a thousandth of its volume, indicating that the proton is 
much more strongly bound.

The relevance to our workshop is that the
polarizability of a hyperon can also be measured using Compton
scattering, via the reaction $B+Z\rightarrow B+Z+\gamma$ extrapolated to
the photon pole---{\it i.e.} the Primakoff effect.   Of course, this is
only feasible for charged hyperons---$\Sigma^\pm,\Xi^-$, and the 
size of such polarizabilities predicted
theoretically are  somewhat smaller than that of the proton\cite{meis}
\begin{equation}
\alpha_E^{\Sigma^+}\sim9.4\times 10^{-4}\,{\rm fm}^3,\qquad
\alpha_E^{\Xi^-}\sim2.1\times 10^{-4}\,{\rm fm}^3
\end{equation}
but their measurement would be of great interest.

\subsection{Polarization and Hyperon Production}

My final topic will be that of polarization in strong interaction
production of hyperons, a field that began here at FNAL in 1976 with
the discovery of $\Lambda$ polarization in the reaction\cite{fnal}
\begin{equation}
p(300\,\,{\rm GeV})+Be\rightarrow\vec{\Lambda}+X
\end{equation}
This process has been well studied in the intervening years\cite{hel} and we
now know that in the fragmentation region the polarization is large and 
negative---$\vec{P}
\cdot\hat{p}_{inc}\times\hat{p}_\Lambda<0$---and that it satisfies
scaling, {\it i.e.} is a function only of $x_F={p^\parallel_\Lambda\over
p_p},p^\perp_\Lambda$ and not of the center of mass energy .  Various 
theoretical approaches have been applied in order to try to understand
this phenomenon---{\it e.g.}, Soffer and T\"{o}rnqvist have developed a
Reggized pion exchange picture\cite{st}, while DeGrand, Markkanen, 
and Miettinen
have used a quark-parton approach wherein the origin of the
polarization is related to  
the Thomas precession\cite{dm}---but none can be said
to be definitive.  One thing which seems to be clear is that there exists
a close connection with the large negative polarizations seen in
inclusive hyperon production and the large positive analyzing powers observed 
at FNAL in inclusive meson production with polarized protons\cite{anp}
\begin{equation}
\vec{p}+p\rightarrow \pi^++X
\end{equation}
Another input to the puzzle may be the availability in the lower
energy region of new exclusive
data from Saturne involving\cite{sat}
\begin{equation}
\vec{p}+p\rightarrow p+\vec{\Lambda}+K^+
\end{equation}
which seems best described in terms of a kaon exchange mechanism.
Clearly there is much more to do in this field.

\section{Summary}  
I conclude by noting that, although the first hyperon was discovered
more then half a century ago and much work has been done since, 
the study of hyperons remains an interesting and challenging
field.  As I have tried to indicate above, many questions exist
as to their strong, weak, and electromagnetic interaction properties,
and I suspect that these particles will remain choice targets for particle
hunters well into the next century.

\end{document}